\documentstyle[epsf]{article}

\newcommand{\newsection}{ \setcounter{equation}{0} \section}
\newcommand{\beq}{\begin{equation}} \newcommand{\eeq}{\end{equation}}
\newcommand{\bea}{\begin{eqnarray}} \newcommand{\eea}{\end{eqnarray}}

  \newcommand
{\Romannumeral}[1]{\uppercase\expandafter{\romannumeral#1}}

\newcommand{\be}{\begin{enumerate}} \newcommand{\ee}{\end{enumerate}}
\newcommand{\bi}{\begin{itemize}} \newcommand{\ei}{\end{itemize}}
\newcommand{\ba}{\begin{array}} \newcommand{\ea}{\end{array}}
\newcommand{\bc}{\begin{center}} \newcommand{\ec}{\end{center}}
\newcommand{\bt}{\begin{tabular}} \newcommand{\et}{\end{tabular}}

\def\lsim{\mathrel{\rlap{\lower4pt\hbox{\hskip1pt$\sim$}}
    \raise1pt\hbox{$<$}}}           
\def\gsim{\mathrel{\rlap{\lower4pt\hbox{\hskip1pt$\sim$}}
    \raise1pt\hbox{$>$}}}           

\newcommand{\half}{\textstyle {1\over2} \displaystyle}    
\newcommand{\quarter}{\textstyle {1\over4} \displaystyle} 
\newcommand{\eigth}{\textstyle {1\over8} \displaystyle}   
\newcommand{\Dslash}{{\hbox{D}\kern-0.6em\raise0.15ex\hbox{/}}} 

\begin{document}

\setlength{\oddsidemargin}{0cm} \setlength{\baselineskip}{7mm}

\input epsf




\begin{normalsize}\begin{flushright}
    DAMTP-97-75 \\
    August 1997 \\
\end{flushright}\end{normalsize}

\begin{center}
  
\vspace{50pt}
  
{\Large \bf ON THE MEASURE IN SIMPLICIAL GRAVITY }

\vspace{40pt}
  
{\sl Herbert W. Hamber}
$^{}$\footnote{e-mail address : hamber@cern.ch; permanent address:
University of California, Irvine Ca 92717 USA}
and {\sl Ruth M. Williams}
$^{}$\footnote{e-mail address : rmw7@damtp.cam.ac.uk; permanent address:
DAMTP, Silver Street, Cambridge CB3 9EW, England}
\\

\vspace{20pt}

Theoretical Physics Division, CERN \\
CH-1211 Geneva 23, Switzerland \\

\end{center}

\vspace{40pt}

\begin{center} {\bf ABSTRACT } \end{center}
\vspace{12pt}
\noindent

Functional measures for lattice quantum gravity should agree with
their continuum counterparts in the weak field, low momentum limit.
After showing that the standard simplicial measure satisfies the above
requirement, we prove that a class of recently proposed non-local
measures for lattice gravity do not satisfy such a criterion, already
to lowest order in the weak field expansion. We argue therefore that
the latter cannot represent acceptable discrete functional measures
for simplicial geometries.

\vspace{24pt}



\vfill

\newpage

\vskip 10pt
\newsection{Introduction}
\hspace*{\parindent}

In the simplicial formulation of quantum gravity one approximates
the functional integration over continuous metrics by a discretized
sum over piecewise linear simplicial geometries.
In such a model the role of the continuum metric is played
by the edge lengths of the simplices, while curvature is naturally
described by a set of deficit angles which can be computed as
functions of the given edge lengths.
It has been known for some time that the simplicial lattice
formulation of gravity is locally gauge invariant, and that it contains
perturbative gravitons in the lattice weak field expansion,
making it an attractive lattice regularization of the continuum theory.

Recent evidence seems to indicate that simplicial quantum gravity
in four dimensions exhibits a phase transition between a smooth
and a rough phase. Only the smooth, small curvature phase appears to
be physically acceptable ~\cite{phases}.
The existence of a phase transitions implies non-trivial and
calculable non-perturbative scaling properties for the coupling
constants of the theory, and in particular Newton's constant.
All calculations so far have been performed in the Euclidean formulation.
As usual, the starting point for a non-perturbative study of quantum gravity
is a suitable definition of the path integral.
In the simplicial lattice approach one starts from the discretized
Euclidean path integral for pure gravity,
with the squared edge lengths as fundamental variables,
\beq
Z_L \; = \; \int_0^\infty \; \prod_s \; \left ( V_d (s) \right )^{\sigma} \;
\prod_{ ij } \, dl_{ij}^2 \; \Theta [l_{ij}^2]  \; 
\exp \left \{ 
- \sum_h \, \Bigl ( \lambda \, V_h - k \, \delta_h A_h 
+ a \, { \delta_h^2 A_h^2 \over V_h } + \cdots \Bigr ) \right \}  \;\; .
\label{eq:zlatt} 
\eeq
The above expression represents a suitable discretization of the
continuum Euclidean path integral for pure quantum gravity
\beq
Z_C \; = \; \int \prod_x \;
\left ( {\textstyle \sqrt{g(x)} \displaystyle} \right )^{\sigma}
\; \prod_{ \mu \ge \nu } \, d g_{ \mu \nu } (x) \;
\exp \left \{ 
- \int d^4 x \, \sqrt g \, \Bigl ( \lambda - { k \over 2 } \, R
+ { a \over 4 } \, R_{\mu\nu\rho\sigma} R^{\mu\nu\rho\sigma}
+ \cdots \Bigr ) \right \}  \;\; ,
\label{eq:zcont}
\eeq
with $k^{-1} = 8 \pi G $.
The $\delta A$ term in the lattice action is the well-known Regge term
~\cite{regge}, which reduces to the Einstein-Hilbert action
in the lattice weak field limit ~\cite{rowi}.
A cosmological constant term is needed for convergence of the path
integral, while the curvature squared term allows one to control the
fluctuations in the curvature.
In the discrete case the integration over metrics is replaced by
integrals over the elementary lattice degrees of freedom,
the squared edge lengths, as discussed in ~\cite{lesh,hartle,gauge}.
The higher derivative terms eventually become irrelevant at distances
much larger than the Planck length, $r \gg \sqrt{a G}$.
For phenomenological reasons one is therefore mostly interested in the
limit $a \rightarrow 0 $, and in this limit the theory depends, in the
absence of matter and after a suitable rescaling of the metric, only on
one bare parameter, the dimensionless coupling $k^2 / \lambda $.

The two phases of quantized gravity found in ~\cite{phases},
can loosely be described as having in one
phase ($G<G_c$, the rough, branched polymer-like phase)
\beq
\langle g_{\mu\nu} \rangle \; = \; 0 \;\; ,
\eeq
while in the other phase ($G>G_c$, the smooth phase),
\beq
\langle g_{\mu\nu} \rangle \; \approx \; c \; \eta_{\mu\nu} \;\; ,
\eeq
with a small negative average
curvature (anti-DeSitter space) in the vicinity of the critical point
at $G_c$, which then vanishes as the critical point is approached from above.
It appears that only the phase $G>G_c$ is physically acceptable,
since in the complementary phase the simplicial lattice degenerates into a
lower-dimensional branched-polymer like manifold, with a proliferation
of sharp curvature singularities, and no physically
acceptable continuum limit.
The challenge of course lies in extracting accurate physical predictions
from the theory as one approaches the lattice continuum limit by
taking $G \rightarrow G_c $  from the smooth, negative curvature phase, side.
It is only in the physical, smooth phase that the simplicial lattice theory
leads to a prediction for the non-perturbative scale dependence
of Newton's constant, which can be cast in the simple form ~\cite{phases}
\beq
G(r) \; = \; G(0) \left [ 1 \, + \, c \, ( r / R_0 )^{1 / \nu} \, 
+ \, O (( r / R_0 )^{2 / \nu} ) \right ] \;\; .
\eeq
Here the critical exponent $1/\nu = 2.8(3) $, and $c$ a numerical
constant of order one;
the scale $R_0^{-1}$ plays a role similar to the scaling violation
parameter $\Lambda_{\overline{MS}}$ in QCD, with $ R_0 \approx c H_0^{-1} $.
A more detailed discussion of the properties of the
two phases characterizing four-dimensional quantum gravity, and of the
computation of the associated critical exponents, can be found
in ~\cite{phases}. 
A description of earlier work on simplicial gravity can be found
in ~\cite{hw84}.
For related work on simplicial gravity
see also the references in ~\cite{monte}, where the same two-phase structure
for four-dimensional simplicial gravity has been observed.
An up-to-date description of work in classical simplicial gravity
and the discrete time evolution problem can be found in ~\cite{miller}.
For results with an alternative and 
complementary approach to problems in quantum gravity
based on dynamical triangulations, we shall point
the reader to the references in ~\cite{smit}.

The functional measure over metrics
is an essential ingredient in the quantum theory of gravity.
In this paper we address the issue of whether the lattice
gravitational measure is unique, and if not how to decide among a set of
different possible lattice measures. 
It is sometimes stated that the universal character of long distance
critical behavior will wash out the difference between similar actions and
measures. While this statement might be true for action terms that
contain higher derivatives, and are therefore potentially irrelevant
in the lattice continuum limit, it is less clear that it applies to the
functional measure.
In this paper we focus on a comparison of different approaches to the
functional measure in simplicial quantum gravity, by examining both
the traditional local measure, as well as highly non-local measures which have
recently been proposed in the literature.
Throughout the paper we shall make use of the fact that in the continuum
the functional measure for quantized gravity is well known and understood.
We then point out the obvious, and natural,
requirement that the lattice functional measure should agree with
the continuum functional measure in the weak field, low momentum limit.
A straightforward lattice perturbative calculation will then show that
this key requirement {\it is} satisfied by
a class of local measures currently used in the numerical simulations,
but that, on the other hand, it is {\it not} satisfied by another set
of non-local measures which have been recently proposed in the literature.
We will conclude therefore that the latter do not represent acceptable
functional measures for simplicial geometries.

\vskip 20pt
\subsection{Standard Measure}
\hspace*{\parindent}

As the edge lengths play the role of the metric in the
continuum, one expects the discrete measure to involve an
integration over the squared edge lengths \cite{lesh,hartle,gauge}.
Indeed the induced metric at a simplex is related to the squared edge
lengths within that simplex, via the expression for the
invariant line element $ds^2 = g_{\mu \nu} dx^\mu dx^\nu$.
After chosing coordinates along the edges emanating from a vertex,
the relation between metric perturbations and squared edge
length variations for a given simplex based at 0 in $d$ dimensions is
\beq
\delta g_{ij} (l^2) \; = \; \half \;
( \delta l_{0i}^2 + \delta l_{0j}^2 - \delta l_{ij}^2 ) \;\; .
\eeq
For one $d$-dimensional simplex labeled by $s$
the integration over the metric is thus equivalent to an 
integration over the edge lengths, and one has
\beq
\left ( {1 \over d ! } \sqrt { \det g_{ij}(s) } \right )^{\sigma} \;
\prod_{ i \geq j } \, d g_{i j} (s) \; = \; 
{\textstyle \left ( - { 1 \over 2 } \right ) \displaystyle}^{ d(d-1) \over 2 }
\; \left [ V_d (l^2) \right ]^{\sigma} \;
\prod_{ k = 1 }^{ d(d+1)/2 } \, dl_{k}^2 \;\; .
\label{eq:simpmeas}
\eeq
There are $d(d+1)/2$ edges for each
simplex, just as there are $d(d+1)/2$ independent components for the metric
tensor in $d$ dimensions.
Here one is ignoring temporarily the triangle inequality constraints,
which will further require all sub-determinants of $g_{ij}$ to be
positive, including the obvious restriction $l_k^2 >0$.
The extension to many simplices glued together at their common faces
is then immediate. For this purpose one first needs to identify edges
$ l_k (s) $ and $ l_{k'} (s') $ which are shared between
simplices $s$ and $s'$,
\beq
\int_0^\infty d l^2_k (s) \, \int_0^\infty d l^2_{k'} (s') \;
\delta \left ( l^2_k (s) - l^2_{k'} (s') \right ) 
\, = \, \int_0^\infty d l^2_k (s) \;\; .
\eeq
After summing over all simplices one derives,
up to an irrelevant numerical constant, the unique functional measure
for simplicial geometries
\beq
\int d \mu [l^2] \; = \; 
\int_0^\infty \; \prod_s \; \left [ V_d (s) \right ]^{\sigma} \;
\prod_{ ij } \, dl_{ij}^2 \; \Theta [l_{ij}^2]  \;\; .
\label{eq:lattmeas}
\eeq
Here $ \Theta [l_{ij}^2] $ is
a (step) function of the edge lengths, with the property
that it is equal to one whenever the triangle inequalities and their
higher dimensional analogs are satisfied,
and zero otherwise.
In four dimensions the lattice analog of the DeWitt measure
($\sigma=0$) takes on a particularly simple form, namely
\beq
\int d \mu [l^2] \; = \; \int_0^\infty \prod_{ ij } \, dl_{ij}^2 
\; \Theta [ l_{ij}^2 ] \;\; .
\label{eq:dewlattmeas}
\eeq
The above lattice measure over the space of squared edge lengths
has been used extensively in numerical simulations of simplicial
quantum gravity ~\cite{phases,lesh,hw84,monte,monte2d,hw2d}.

The derivation of the above lattice measure
closely parallels the analogous procedure in the
continuum. There, following DeWitt ~\cite{dewitt,fuji},
one defines an invariant norm for metric fluctuations 
\beq
\Vert \delta g \Vert^2 \; = \;
\int d^d x \left ( g(x) \right )^{\omega/2} \;
G^{\mu \nu, \alpha \beta} [g(x); \omega ] \;
\delta g_{\mu \nu}(x) \, \delta g_{\alpha \beta}(x) \;\; ,
\eeq
with the inverse of the super-metric $G$ given by
\beq
G^{\mu \nu, \alpha \beta} [g(x); \omega] \; = \;
\half \; \left ( g(x) \right )^{(1-\omega)/2} \left [
g^{\mu \alpha}(x) g^{\nu \beta}(x) +
g^{\mu \beta}(x) g^{\nu \alpha}(x) + \lambda \,
g^{\mu \nu}(x) g^{\alpha \beta}(x) \right ] \;\; .
\label{eq:dewittsuper}
\eeq
DeWitt originally considered the case $\omega=0$, but it will be
useful later to consider other values for $\omega$, such as $\omega=1$.
The resulting functional measure in the continuum is then
given by
\beq
\int d \mu [g] \; = \; \int \prod_x \left [ \det G(g(x)) \right ]^{\half} 
\prod_{\mu \geq \nu} d g_{\mu \nu} (x) \;\; .
\eeq
Since
the super-metric $G^{\mu \nu, \alpha \beta} (g(x))$ is ultra-local, one
expects its determinant to be a local function of $x$ as well.
Up to an irrelevant multiplicative constant, one has for the
determinant of $G$ the simple result
\beq
\det G (g(x)) \propto (1 + \half d \lambda ) \;
[ g(x) ]^{ (d+1)((1- \omega )d-4)/4 }
\;\; .
\label{eq:dewittdet}
\eeq
One also needs to impose the condition $\lambda \neq - 2 / d $
in order to avoid the vanishing of the determinant of $G$.
As a result, one obtains the local measure for the functional
integration over metrics
\beq
\int d \mu [ g] \; = \; \int \prod_x \;
\left [{\textstyle \sqrt{g(x)} \displaystyle} \right ]^{\sigma}
\; \prod_{ \mu \ge \nu } \, d g_{ \mu \nu } (x) \;\; ,
\label{eq:contmeas}
\eeq
with $\sigma \, = \, (d+1)[(1- \omega )d-4] / 4 $.
For $\omega=0$ one obtains the DeWitt measure for pure
gravity, which takes on a particularly simple form
in $d=4$,
\beq
\int \prod_x \; [ g(x) ]^{ (d-4)(d+1)/8 } \;
\prod_{\mu \ge \nu} \, d g_{\mu \nu} (x)
\; \mathrel{\mathop\rightarrow_{ d = 4}} \;
\int \prod_x \prod_{\mu \geq \nu} d g_{\mu \nu} (x) \;\; , 
\label{eq:dewitt}
\eeq
and which obviously corresponds to the lattice measure in
Eq.~(\ref{eq:dewlattmeas}). In general the volume factors are
absent ($\sigma=0$) if one choses $\omega = { d-4 \over d}$.
On the other hand, for $\omega=1$ one recovers the Misner
measure ~\cite{misner,fadeev}.
\footnote{
It is easy to show that the continuum measure
of Eq.~(\ref{eq:contmeas}) is invariant
under coordinate transformations, irrespective of the value of
$\sigma$.
Under a change of coordinates ${x'}^{\mu} = x^{\mu} + \epsilon^{\mu} (x)$
\beq
\prod_x \;
\left [ g(x) \right ]^{\sigma / 2}
\; \prod_{ \mu \ge \nu } \, d g_{ \mu \nu } (x)
\; \mathrel{\mathop\rightarrow} \;
\prod_x \;
\left ( \det { \partial {x'}^{\beta} \over \partial {x}^{\alpha} }
\right )^{\gamma} \;
\left [ g(x) \right ]^{\sigma / 2}
\; \prod_{ \mu \ge \nu } \, d g_{ \mu \nu } (x) \;\; .
\eeq
For infinitesimal coordinate transformations the additional factor
is equal to one,
\beq
\prod_x \;
\left ( \det { \partial {x'}^{\beta} \over \partial {x}^{\alpha} }
\right )^{\gamma} \;
\; = \; 
\prod_x \;
\left [ \det ( \delta_{\alpha}^{\;\; \beta} + \partial_{\alpha}
\epsilon^{\beta} ) \right ]^{\gamma} \;
\; = \; 
\exp \left \{ \gamma \, \delta^d (0) \int d^d x \; \partial_{\alpha}
\epsilon^{\alpha} \right \} \; = \; 1 \;\; .
\eeq
In many respects $\sigma$ can be thought of as a gauge parameter.
}

There is no clear way of deciding between these two choices
($\omega=0$ or 1), or any
intermediate one for that matter, and one should consider
$\sigma$ as an arbitrary parameter of the model, to be
constrained only by the requirement that the path integral be well defined
(which incidentally rules out singular measures).
Note that the volume term in the measure is 
completely local and contains no derivatives.
In perturbation theory it does not therefore effect the propagation
properties of
gravitons, and contributes $\delta^d(0)$ terms to the effective
action; to some extent these can be regarded as similar to
a renormalization of the cosmological constant, affecting only
the distribution of local volumes.
Numerical simulations in the lattice model show very little
sensitivity of the critical exponents to either $\sigma$ or $a$
~\cite{phases}.

There is no obstacle in defining a discrete analog of the
supermetric, as a way of introducing an invariant notion of distance
between simplicial manifolds.
It leads to an alternative way of deriving the lattice measure in
Eq.~(\ref{eq:dewlattmeas}), by
considering the discretized distance between induced metrics
$g_{ij}(s)$ ~\cite{cms},
\beq
\Vert \, \delta g (s) \, \Vert^2 \; = \; \sum_{s} \;
G^{ i j k l } \left ( g(s) \right ) \; 
\delta g_{i j} (s) \, \delta g_{k l} (s) \;\; ,
\eeq
with the inverse of the lattice DeWitt supermetric now given by
the expression
\beq
G^{ i j k l } [ g(s) ] \; = \; 
\half \sqrt{g(s)} \; \left [ \,
g^{i k} (s) g^{j l} (s) +
g^{i l} (s) g^{j k} (s) + \lambda \,
g^{i j} (s) g^{k l} (s) \right ] \;\; ,
\label{eq:dewittsuperl}
\eeq
and with again $\lambda \neq - 2 / d $.
This procedure defines a metric on the tangent space of positive real
symmetric matrices $g_{ij}(s)$. After computing the determinant
of $G$, the resulting functional measure is
\beq
\int d \mu [l^2] \; = \; \int
\prod_{s} \, \left [ \; \det G(g(s)) \; \right ]^{\half}
\prod_{i \geq j} d g_{i j} (s) \;\; ,
\eeq
with the determinant of the super-metric $G^{i j k l} (g(s))$
given by the local expression
\beq
\det G (g(s)) \; \propto \; 
(1 + \half d \lambda ) \; \left [ g(s) \right ]^{ (d-4)(d+1)/4 } \;\; ,
\eeq
Using Eq.~(\ref{eq:simpmeas}). and up to irrelevant constants,
one obtains again the standard lattice measure
of Eq.~(\ref{eq:lattmeas}). 
Of course the same procedure can be followed for the Misner-like
measure, leading to a similar result for the lattice measure,
but with a different power $\sigma$. For a related discussion see
also ~\cite{ban}.

\vskip 20pt
\subsection{Alternative Approach}
\hspace*{\parindent}

The previous derivation of the standard lattice functional measure is
based on the direct and obvious correspondence between the induced
lattice metric within a simplex and the continuum metric at a point.
It leads to an essentially unique local measure over the
squared edge lengths, in close analogy to the continuum expression.
In particular it is clear from the derivation that the lattice and continuum
measures  agree with each other in the weak field expansion,
essentially by construction.

Still, one might be tempted to try to find an alternative lattice
measure by looking directly at the discrete form for the supermetric,
written as a quadratic form in the squared edge lengths (instead of
the metric components), and then evaluating the resulting determinant.
The main idea, inspired by work described in an unpublished paper
by Lund and Regge ~\cite{lund} on the $3+1$ formulation of simplicial
gravity, can be found in some detail in a recent paper ~\cite{gauge};
see also another recent paper ~\cite{hmw}, which discusses somewhat
different issues, not directly related to the measure.
First one considers a lattice analog of the DeWitt supermetric,
by writing
\beq
\Vert \delta l^2 \Vert^2 \; = \; \sum_{ij} \; G_{ij} (l^2)
\; \delta l^2_i \; \delta l^2_j \; \; ,
\label{eq:lund}
\eeq
with $G_{ij} (l^2)$ playing a role analogous to the DeWitt supermetric,
but defined now on the space of squared edge lengths.
The next step is to find an appropriate form for $G_{ij} (l^2)$ expressed in
terms of known geometric objects.
One simple way of constructing the explicit form for $G_{ij} (l^2)$, in any
dimension, is to first focus on one simplex, and
write the squared volume of a given simplex in terms
of the induced metric components within the {\it same} simplex $s$,
\beq
V^2 ( s ) \; = \; {\textstyle \left ( { 1 \over d! } \right )^2 \displaystyle}
\det g_{ij}(l^2(s)) \;\; .
\eeq
One computes to linear order
\beq
{1 \over V (l^2)} \; \sum_{i} 
{\partial V^2 (l^2) \over \partial l^2_i} \; \delta l^2_i
\; = \; {\textstyle { 1 \over d! } \displaystyle} \sqrt{ \det ( g_{ij} ) }
\; g^{ij} \; \delta g_{ij} \;\; ,
\eeq
and to quadratic order
\beq
{1 \over V (l^2)} \; 
\sum_{ij} { \partial^2 V^2 (l^2) \over \partial l^2_i \partial l^2_j }
\; \delta l^2_i \; \delta l^2_j \; = \;
{\textstyle { 1 \over d! } \displaystyle} \sqrt{ \det ( g_{ij} ) }
\left [ \; g^{ij} g^{kl} \delta g_{ij} \delta g_{kl}
- g^{ij} g^{kl} \delta g_{jk} \delta g_{li} \; \right ] \;\; .
\eeq
The right hand side of this equation contains precisely
the expression appearing in the continuum supermetric
of Eq.~(\ref{eq:dewittsuper}), for the specific choice of the
parameter $\lambda = -2$.
One is lead therefore to the obvious identification
\beq
G_{ij} (l^2) \; = \; - \; d! \; \sum_{s} \;
{1 \over V (s)} \; 
{ \partial^2 \; V^2 (s) \over \partial l^2_i \; \partial l^2_j } \;\; ,
\eeq
and therefore for the norm
\beq
\Vert \delta l^2 \Vert^2 \; = \; \sum_{s} \; V (s) \;
\left \{
\; - \; {d! \over V^2 (s)} \; \sum_{ij} \;
{ \partial^2 \; V^2 (s) \over \partial l^2_i \; \partial l^2_j }
\; \delta l^2_i \; \delta l^2_j \; \right \} \;\; .
\eeq
One could be tempted at this point to write down a lattice measure,
in parallel with Eq.~(\ref{eq:dewittsuper}), and write
\beq
\int \; d \mu [ l^2 ] \; = \; \int \; \prod_i \;
\sqrt{ \det G_{ij}^{( \omega ' )} (l^2) } \, dl^2_i \;\; .
\label{eq:lundmeas}
\eeq
with 
\beq
G^{(\omega')}_{ij} (l^2) \; = \; - \; d! \; \sum_{s} \;
{ 1 \over [ V(s) ]^{1+{\omega'}} } \;
{ \partial^2 \; V^2 (s) \over \partial l^2_i \; \partial l^2_j } \;\; ,
\label{eq:gmatrix}
\eeq
Again we have allowed here for a parameter $\omega'$, which is possibly
different from zero, and interpolates between apparently equally
acceptable measures.
As in the continuum, different edge length measures, here parametrized by
$\omega$', are obtained, depending on whether the local volume
factor $V(s)$ is included in the supermetric or not.
Irrespective of the value chosen for $\omega'$, we will show
below that the measure of Eq.~(\ref{eq:lundmeas}) disagrees with
the continuum measure of Eq.~(\ref{eq:contmeas}) already to lowest
order in the weak field expansion,
and does not therefore describe an acceptable lattice measure.

An obvious undesirable (and puzzling) feature of the measure of 
Eq.~(\ref{eq:lundmeas}) is that in 
general it is non-local, in spite of the fact that the original
continuum measure of Eq.~(\ref{eq:contmeas}) {\it is} completely local
(although it is clear that for some special choices
of $\omega'$ and $d$, one {\it does} recover a local measure;
thus in two dimensions and for ${\omega'} = -1$ one obtains again
the simple result
$\int \; d \mu [ l^2 ] \; = \; \int_0^\infty \; \prod_i \, dl^2_i \;$).
It was already pointed out in ~\cite{gauge} that the above 
procedure also fails to give the correct measure already
in one dimension.

Let us now turn to the calculation of the determinant $\det G(l^2) $.
In general it is given by a rather formidable expression, which can
be simplified though by considering its lattice weak field expansion,
and which will allow us to make a direct comparison with the
continuum answer of Eq.~(\ref{eq:dewittdet}).
In order to discuss the weak field expansion of the lattice
measure of Eq.~(\ref{eq:lundmeas}), we shall focus here for simplicity
on the two-dimensional case, for which an explicit answer can readily
be obtained; although our arguments are general,
the algebraic complexity is significantly reduced in two
dimensions. Also for definiteness we will consider the case
$\omega'$=0 in Eq.~(\ref{eq:gmatrix}).
It is clear that the determinant, being a non-local
function of the edge lengths, will couple edges which are 
arbitrarily far apart on the lattice.
For a square lattice made rigid by the introduction of diagonals,
$ G(l^2) $ will be an $3N_0 \times 3N_0$ matrix, with $N_0$ denoting the
total number of sites in the lattices. 
It will be sufficient in the following to examine the form of $\det G (l^2)$ 
for a square lattice with 12 edges (see Figure 1.), with the usual
imposition of periodic boundary conditions to minimize edge effects.

\vskip 90pt

\begin{center}
\leavevmode
\epsfysize=7cm
\epsfbox{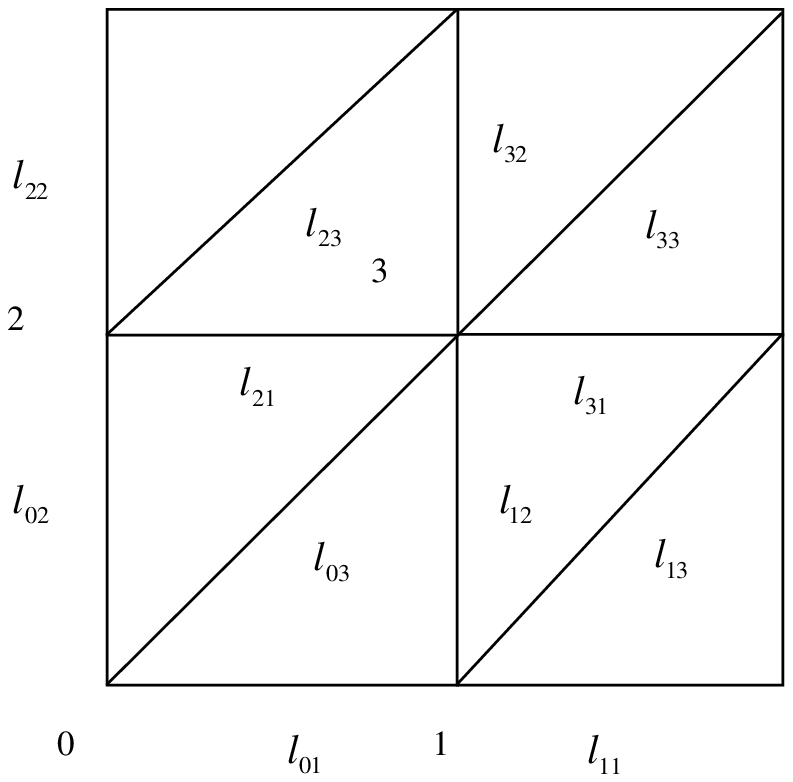}
\end{center}
\noindent
{\small{\it Fig.\ 1.
Notation for the weak-field expansion about the rigid square lattice.
\medskip}}

For such a lattice $G(l^2)$ is given by the symmetric $12 \times 12$ matrix
\beq
G (l^2) \; = \; {1 \over 4} \left( \begin{array}{ccccccc}

         {1\over A_{01}}+{1 \over A_{22}}&0&-{1\over A_{01}}&
         0&-{1\over A_{01}}&0& \cdots \\
         0&{1\over A_{02}}+{1 \over A_{11}}&-{1\over A_{02}}& -{1
         \over A_{11}}&0&-{1 \over A_{11}}& \cdots \\
         -{1\over A_{01}}&-{1\over A_{02}}&{1\over A_{01}}+{1\over
         A_{02}}& 0&-{1\over A_{01}}&0& \cdots \\

         0&-{1 \over A_{11}}&0& {1 \over A_{11}}+{1 \over
         A_{32}}&0&-{1 \over A_{11}}& \cdots \\ 
         -{1\over A_{01}}&0&-{1\over A_{01}}& 0&{1\over A_{01}}+{1
         \over A_{12}}&-{1 \over A_{12}}& \cdots \\
         0&-{1 \over A_{11}}&0& -{1 \over A_{11}}&-{1 \over A_{12}}&{1
         \over A_{11}}+{1 \over A_{12}}& \cdots \\

         0&-{1\over A_{02}}&-{1\over A_{02}}& 0&0&0& \cdots \\
         -{1 \over A_{22}}&0&0& 0&0&0& \cdots \\
         -{1 \over A_{22}}&0&0& 0&0&0& \cdots \\

         0&0&0& 0&-{1 \over A_{12}}&-{1 \over A_{12}}& \cdots \\
         0&0&0& -{1 \over A_{32}}&0&0& \cdots \\
         0&0&0& -{1 \over A_{32}}&0&0& \cdots \\

\end{array} \right)
\eeq
where $A_{i1}$ and $A_{i2}$ denote the areas of the two triangles
based at site $i$.
The area of a triangle with arbitrary edge lengths $l_1$,  $l_2$, and
$l_3$ is given here as usual in terms of the edge lengths by
\beq
A_T (l_1,l_2,l_3) \; = \; {1 \over 4} \sqrt{ 2 (l_1^2 l_2^2 + 
l_2^2 l_3^2 + l_3^2 l_1^2) - l_1^4 - l_2^4 - l_3^4 } \;\; .
\eeq
After expanding it in terms of the edge lengths, the determinant
$\det G(l^2)$ is in the general case given by a rather complicated expression.
To make progress, one can further expand it for small fluctuations
in the edge lengths.
It is convenient for this purpose to use a binary notation
~\cite{rowi} for the vertices, and introduce small edge length fluctuations
$\epsilon_i$, by writing
\beq
l_i \; = \; l_i^0 \; (1 + \epsilon_i) \;\; ,
\label{eq:epsilon} 
\eeq
with $l_1^0 = l_2^0 =1$ and $l_3^0 = \sqrt{2}$ for a square background 
lattice (see again Figure 1.).
The individual triangle areas can in turn be expanded in term of the
$\epsilon$'s, to give for example
\beq 
A_{01} (\epsilon) \; = \; { \half } +
{ \half } (\epsilon_{01} + \epsilon_{12} ) +
{ \quarter } (\epsilon_{01} \epsilon_{03} + \epsilon_{03} \epsilon_{12}
- \epsilon_{01}^2 - \epsilon_{12}^2 - 4 \epsilon_{03}^2 )
+ O( \epsilon^3 )
\eeq
and similarly for the remaining triangle areas.
Our notation here is that the first index labels the site and the second
one the lattice direction.
It can be shown that the expansion needs to be carried out to fourth
order in $\epsilon$ in order to get a non-vanishing result for the
determinant of $G(l^2)$.
The resulting expressions are then inserted into the formula
for the determinant and give, for the square lattice,
\bea
\det G( \epsilon ) \; = \; & { 1 \over 2 } \;
(\epsilon_{01} + \epsilon_{11} - \epsilon_{21} - \epsilon_{31})
\; (\epsilon_{02} - \epsilon_{12} + \epsilon_{22} - \epsilon_{32}) 
\nonumber \\ 
& \times (2 \epsilon_{01} \epsilon_{03} + 2 \epsilon_{02} \epsilon_{03} -
4 \epsilon_{03}^2 + \epsilon_{02} \epsilon_{11} -
\epsilon_{01} \epsilon_{12} + 2 \epsilon_{03} \epsilon_{12}
- 2 \epsilon_{02} \epsilon_{13}
\nonumber \\ 
& - 2 \epsilon_{11} \epsilon_{13} -
2 \epsilon_{12} \epsilon_{13} + 4 \epsilon_{13}^2 - 
\epsilon_{02} \epsilon_{21} + 2 \epsilon_{03} \epsilon_{21} +
\epsilon_{01} \epsilon_{22} - 2 \epsilon_{01} \epsilon_{23}
\nonumber \\ 
& - 2 \epsilon_{21} \epsilon_{23} - 2 \epsilon_{22} \epsilon_{23} + 
4 \epsilon_{23}^2 + \epsilon_{12} \epsilon_{31} -
2 \epsilon_{13} \epsilon_{31} - \epsilon_{22} \epsilon_{31} - 
\epsilon_{11} \epsilon_{32}
\nonumber \\ 
& + \epsilon_{21} \epsilon_{32} -
2 \epsilon_{23} \epsilon_{32} + 2 \epsilon_{11} \epsilon_{33} +
2 \epsilon_{22} \epsilon_{33} + 2 \epsilon_{31} \epsilon_{33} + 
2 \epsilon_{32} \epsilon_{33} - 4 \epsilon_{33}^2 )
+ O( \epsilon^5 )  \;\; .
\nonumber \\ 
\label{eq:det_e}
\eea
As expected, the result is indeed non-local, and
involves to this order contributions from all the edges on the
4-site lattice. It is in fact easy to see that this will be the case
for any size lattice, due to the general non-locality of the determinant.
As a check of the calculation, one can verify that as the $\epsilon$'s
approach zero, one recovers the zero eigenvalues of the matrix $G$
for the square lattice, with the correct multiplicity
(the eigenvalues for $G$ in this case are
$-1, 3 \times 0, 3 \times 1, 5 \times 2$).
A somewhat simpler and more symmetric expression is obtained in the case
of an equilateral lattice, for which one can show that
\beq
\det G( \epsilon ) \; = \; { 2^{15} \over 3^9 } \;
(\epsilon_{01} + \epsilon_{11} - \epsilon_{21} - \epsilon_{31})
\; (\epsilon_{02} - \epsilon_{12} + \epsilon_{22} - \epsilon_{32}) 
\; (\epsilon_{03} - \epsilon_{13} - \epsilon_{23} + \epsilon_{33})
+ O( \epsilon^4 )  \;\; ,
\label{eq:det_e_eq}
\eeq
reflecting the permutation symmetry under the interchange of the
three coordinate directions in this case. Note also that for this
choice of background lattice the 
determinant is now of cubic order in the $\epsilon$'s.
In this case one can verify again that, as the $\epsilon$'s approach zero,
one recovers correctly the three zero eigenvalues of the matrix $G$ for the
equilateral lattice.

The above expression for the determinant on the square lattice
case can be simplified a bit by going to momentum space.
Here we shall take the $\epsilon_i$'s to be plane waves.
When transforming to momentum space, 
one assumes that the fluctuation $\epsilon_i$ 
at the point $i$, $j$ steps in one coordinate direction and $k$ steps in the 
other coordinate direction from the origin, is related to the 
corresponding $\epsilon_i$ at the origin by 
\beq
\epsilon_i^{(j+k)} \, = \, \omega_1^j \, \omega_2^k \, \epsilon_i^{(0)} \;\; ,
\eeq
where $\omega_i=e^{- i k_{i} }$ and $k_i$ is the momentum in
the direction $i$.
Inserting the above expression into the weak-field expression for the
determinant, Eq.~(\ref{eq:det_e}), one obtains (still in the weak
field limit)
\beq
\det G (\epsilon) \; = \; 
( e^{i k_1} - 1)^2 ( e^{i k_1} + 1)^2
( e^{i k_2} - 1)^2 ( e^{i k_2} + 1)^2 \,
\epsilon_{1}^{(0)} (k) \epsilon_{2}^{(0)} (k) \epsilon_{3}^{(0)} (k) \;
[ \epsilon_{1}^{(0)} (k) + \epsilon_{2}^{(0)} (k) - 2
\epsilon_{3}^{(0)} (k) ] \;\; ,
\label{eq:det_o}
\eeq
which can formally be expanded for small momenta to give
\beq
\det G (\epsilon) \; = \;  2^4 \, 
\epsilon_{1}^{(0)} (k) \epsilon_{2}^{(0)} (k) \epsilon_{3}^{(0)} (k) \;
[ \epsilon_{1}^{(0)} (k) + \epsilon_{2}^{(0)} (k) - 2
\epsilon_{3}^{(0)} (k) ] \;
k_1^2 k_2^2 + O(k^5)   \;\; .
\label{eq:det_k}
\eeq
If the lattice periodicity is imposed on the momenta, then the expression
in Eq.~(\ref{eq:det_o}) vanishes identically for plane waves,
while in general Eq.~(\ref{eq:det_e}) does not.
\begin{center}
\leavevmode
\epsfysize=5cm
\epsfbox{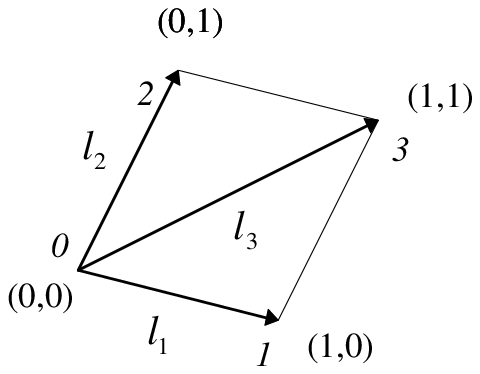}
\end{center}
\noindent
{\small{\it Fig.\ 2.
Edge lengths and metric components.
\medskip}}

The above expression for the determinant can be transformed into
an equivalent form involving the metric field, using the fact
that the edge lengths on the lattice correspond to the metric degrees of
freedom in the continuum. 
Given the choice of edges in Figure 2, one writes for the induced
metric at the origin
\beq
g_{ij} (l^2) \; = \; \left( \begin{array}{cc}
l_1^2 & \half (l_3^2 - l_1^2 - l_2^2 ) \\
\half (l_3^2 - l_1^2 - l_2^2 ) & l_2^2 \\
\end{array} \right) \;\; .
\label{eq:gij_square}
\eeq
One can then relate the edge lengths $l_i$ (or, equivalently, the
fluctuations $\epsilon_i$) to the metric components in the continuum,
which in the weak field limit are more conveniently written as
\beq
g_{ \mu \nu } \; = \; \delta_{ \mu \nu } \; + \; h_{ \mu \nu } \;\; .
\eeq
One then obtains the obvious correspondence between squared edge
lengths and metric components at each lattice vertex
\bea
l_1^2 \; = \; ( 1 + \epsilon_1 )^2 & \, = \, & 1 + h_{11}
\nonumber \\ 
l_2^2 \; = \; ( 1 + \epsilon_2 )^2 & \, = \, & 1 + h_{22}
\nonumber \\ 
\half l_3^2 \; = \; ( 1 + \epsilon_3 )^2 & \, = \, & 1 +
\half ( h_{11} + h_{22} ) + h_{12}  \;\; ,
\nonumber \\ 
\label{eq:htoeps}
\eea
which can be inverted to give the small edge length fluctuations in
terms of the metric components
\bea
\epsilon_1 (h) & \, = \, & \half h_{11} - \eigth h_{11}^2 + O(h_{11}^3)
\nonumber \\ 
\epsilon_2 (h) & \, = \, & \half h_{22} - \eigth h_{22}^2 + O(h_{22}^3)
\nonumber \\ 
\epsilon_3 (h) & \, = \, & \quarter ( h_{11} + h_{22} + 2 h_{12} )
- {\textstyle {1\over32} \displaystyle} 
( h_{11} + h_{22} + 2 h_{12} )^2 + O(h^3)
\nonumber \\ 
\label{eq:epstoh}
\eea
at each point. It is also known that
this relationship is the correct one for relating
edge lengths and continuum metric components in the weak field expansion
for the lattice action, as shown in detail in reference ~\cite{hw2d,gauge}.
Inserting then these expressions into the weak-field
lattice formula for the determinant of Eq.~(\ref{eq:det_k})
one obtains
\beq
\det(G(h)) \; = \;
- \, h_{11}(k) h_{12}(k) h_{22}(k)
\, [ h_{11}(k) + 2 h_{12}(k) + h_{22}(k) ]
\, k_1^2 k_2^2 + O(k^5)
\label{eq:det_h}
\eeq
At this point, one is ready to compare the resulting expression
for the lattice functional measure to the continuum result,
as given in Eq.~(\ref{eq:contmeas}).
In the continuum case one has, in the weak field expansion,
\beq
\det g(x) \, = \, 1 + h_{11}(x) + h_{22}(x) +
h_{11}(x) h_{22}(x) - h_{12}^2(x) + O (h^3)
\eeq
and therefore the functional measure is given by (see Eq.~(\ref{eq:contmeas}))
\beq
\int d \mu [ g] \; = \; \int \prod_x \;
\left ( 1 + h_{11}(x) + h_{22}(x) + \cdots \right )^{\sigma \over 2}
\; \prod_{ \mu \ge \nu } \, d h_{ \mu \nu } (x) \;\; .
\eeq
On the simplicial lattice this last expression obviously becomes
\beq
\; = \; 2^{3 N_0} \; \int \prod_{n=1}^{N_0} \;
\left (  1 + 2 \epsilon_1^{(n)} + 2 \epsilon_2^{(n)}
+ \cdots \right )^{\sigma \over 2}
\; \prod_{i=1}^3 \, d \epsilon_i^{(n)} \;\; .
\eeq
which is clearly very different from the measure of
Eq.~(\ref{eq:lundmeas}), with the determinant $\det G$ given 
(for $\omega'=0$) either by the general weak-field answer of
Eq.~(\ref{eq:det_e}) or, for plane waves, by Eqs.~(\ref{eq:det_o}) and
(\ref{eq:det_h}).

One concludes therefore that the nonlocal measure of
Eq.~(\ref{eq:lundmeas}), which was proposed in ~\cite{nonl}
as a ``new'' measure for simplicial gravity, disagrees with
the continuum measure already to leading order in the weak field
expansion.

\vspace{20pt}
\newsection{Conclusions}
\hspace*{\parindent}

In this paper we have compared different approaches to the functional 
measure in simplicial quantum gravity. We have pointed out that the obvious
requirement that the lattice measure agree with the continuum measure
in the weak field, low momentum limit is satisfied by a class of local
measures used extensively for numerical simulations.
We have also shown that the same requirement is not satisfied by
another set of non-local measures.
The latter do not therefore in our opinion represent acceptable
functional measures for simplicial geometries.
In general we believe that the criterion that lattice operators should
agree with their continuum counterparts in the weak field, low
momentum limit is an important one, and that it should be checked
systematically for any proposed variant action or measure.

\vspace{20pt}

{\bf Acknowledgements}

The authors thank G. Veneziano and the Theory Division at CERN
for hospitality during the completion of this work.
The work of RMW was supported in part by PPARC.

\vfill
\newpage
\end{document}